\documentclass[prl,twocolumn,superscriptaddress,showpacs]{revtex4}

\usepackage{graphicx}
\usepackage{bm}

\newcommand{\beq}{\begin{equation}}
\newcommand{\eeq}{\end{equation}}
\newcommand{\beqa}{\begin{eqnarray}}
\newcommand{\eeqa}{\end{eqnarray}}

\newcommand{\natc}{\emph{i}-NC$_{60}$}

\begin{document}

\title{High Fidelity Single Qubit Operations using Pulsed EPR}

\author{John~J.~L.~Morton}
\email{john.morton@materials.ox.ac.uk} \affiliation{Department of
Materials, Oxford University, Oxford OX1 3PH, United Kingdom}

\author{Alexei~M.~Tyryshkin}
\affiliation{Department of Electrical Engineering, Princeton
University, Princeton, NJ 08544, USA}

\author{Arzhang~Ardavan}
\affiliation{Clarendon Laboratory,
Department of Physics, Oxford University, Oxford OX1 3PU, United
Kingdom}

\author{Kyriakos~Porfyrakis}
\affiliation{Department of Materials, Oxford University, Oxford
OX1 3PH, United Kingdom}

\author{S.~A.~Lyon}
\affiliation{Department of Electrical Engineering, Princeton
University, Princeton, NJ 08544, USA}

\author{G.~Andrew~D.~Briggs}
\affiliation{Department of Materials, Oxford University, Oxford
OX1 3PH, United Kingdom}

\date{\today}

\begin{abstract}
Systematic errors in spin rotation operations using simple RF pulses place severe limitations on the
usefulness of the pulsed magnetic resonance methods in quantum computing applications. In particular, the fidelity of quantum logic operations performed on electron spin qubits falls well below the threshold for the application of quantum algorithms. Using three independent techniques, we demonstrate the use of composite pulses to improve this fidelity by several orders of magnitude. The observed high-fidelity operations are limited by pulse phase errors, but nevertheless fall within the limits required for the application of quantum error correction.
\end{abstract}

\pacs{03.67.Lx, 76.30.-v, 81.05.Tp}

\maketitle


Pulsed magnetic resonance, which provides a way of manipulating quantum systems such as nuclear and electron spins, has proved to be a powerful tool in the development of quantum computation. This is borne out by the early success of nuclear magnetic resonance (NMR)
quantum computers, which have performed the largest-scale quantum
computations to date~\cite{nielsen00,firstNMRQC}. Always-on
dipolar or exchange interactions are exploited to yield
multi-qubit gates, whilst single-qubit operations are performed
using classical radio frequency (RF) pulses. The scalability
limitations surrounding NMR implementations~\cite{warren1997} that arise from the small nuclear Zeeman energy can
be overcome by turning to \emph{electron} paramagnetic resonance
(EPR), analogous in many ways to NMR but with the advantage that
pure states are experimentally accessible.

These strengths have prompted many EPR-based solid state quantum
information processing (QIP)
proposals~\cite{burkard00,harneit,briggsRS,lyon}.  The merit of such 
schemes is often argued on the basis of the decoherence time $T_2$.
However, the fidelity with
which operations can be performed also imposes severe limitations on the viability of such proposals. 
We recently developed a methodology for characterising systematic errors
in pulsed EPR, and used it to measure
typical errors in a commercial EPR spectrometer~\cite{morton04}.
The most significant error present in a single-qubit operation is in the rotation angle, arising from spatial
inhomogeneity in the pulsed RF field. This systematic error in
rotation angle is likely to persist even in
the case of single-molecule EPR experiments, caused by miscalibrated
control equipment. Fortunately, a number of approaches to tackling
different classes of systematic errors have been developed in the
art of NMR, employing composite rotation
sequences~\cite{spinchoreo,Levitt1986}. A small subset of such
approaches correct rotation operators rather than final states, and are therefore successful regardless of the initial spin state. These approaches are applicable to quantum computing~\cite{wimperis,jones03,chuang04,collin04}.  Of these, the BB1 sequence~\cite{wimperis} exploits the precision in pulse phase control to correct for systematic errors in rotation angle, and is therefore ideal for our purposes.

High-fidelity pulses are also beneficial to more traditional EPR
characterisation techniques such as correlation spectroscopy (e.g.
2D-HYSCORE~\cite{schweiger}) by eliminating or suppressing any spurious cross-peak
signals and thus simplifying the spectral analysis.

In this Letter we show how composite pulses can be applied in
pulsed EPR to perform high-fidelity operations on electron spins. We use these to demonstrate non-decaying Rabi
oscillations, and provide further evidence using an
error-sensitive electron spin echo envelope modulation (ESEEM)
effect~\cite{eseem03}. Finally, we use the error-measuring
sequences described in~\cite{morton04} to estimate the residual
error in the composite pulse. We find that in current EPR spectrometers, the chief
limitation lies in the ability to accurately set the pulse phase, which restricts
the effectiveness of the composite pulse.

The paramagnetic species used here to perform error measurements
in pulsed EPR is \emph{i}-NC$_{60}$ (also known as N@C$_{60}$),
consisting of an isolated nitrogen atom in the $^4$S$_{3/2}$
electronic state incarcerated by a C$_{60}$ fullerene cage. It is
an ideal system for these measurements because of its extremely
narrow EPR linewidth and long relaxation time in liquid
solution~\cite{Dietel99,Knapp1997}. $T_{2}$ has been measured to
be $80~\mu$s at room temperature, rising to $240~\mu$s at
170~K~\cite{eseem03}.

The production and subsequent purification of \natc~is described
elsewhere~\cite{mito}. High-purity \emph{i}-NC$_{60}$ powder was
dissolved in CS$_{2}$ to a final concentration of
10$^{15}$/cm$^3$, freeze-pumped to remove oxygen, and finally
sealed in a quartz EPR tube. Samples were 0.7-1.4~cm long, and
contained approximately $5\cdot 10^{13}$ \emph{i}-NC$_{60}$ molecules.
Pulsed EPR measurements were made at 190~K using an X-band Bruker
Elexsys580e spectrometer, equipped with a nitrogen-flow cryostat.

\emph{i}-NC$_{60}$ has electron spin $S=3/2$ coupled to the
$^{14}$N nuclear spin $I=1$. The EPR spectrum consists of three
lines centered at electron g-factor $g=2.003$ and split by a
$^{14}$N \emph{isotropic} hyperfine interaction $a=0.56$~mT in
CS$_2$~\cite{Murphy1996}. Most of the pulsed EPR experiments discussed
below were performed using selective pulses on the central hyperfine line in the EPR triplet,
corresponding to $^{14}$N nuclear spin projection $M_I=0$, for which we can use a vector representation in
visualising the evolution under RF pulses~\cite{schweiger,morton04}. The ESEEM experiment was performed on the $M_I=-1$ hyperfine line, for which a full spin density matrix treatment is necessary~\cite{eseem03}.

In order to measure the quality of an operation independently of the starting state,
we define the fidelity, ${\cal F}$,  which compares the operator for the actual rotation with that of the ideal rotation. The fidelity, 
\beq {\cal F} = \frac{1}{2}{\rm Tr}\left(A B^{-1}\right)
\label{fideq}\eeq
takes a value between 0 and 1 depending on how well the composite rotation $B$ approximates the ideal rotation $A$ (where $A$ and $B$ are unitary matrix operators).

A general rotation of desired angle $\theta$ with systematic error
$\epsilon$, about an in-plane axis $\phi$ is given by
\beq {\cal R}_{\phi}[\theta(1+\epsilon)] = e^{i \left( \sigma_x\cos{\left(\phi\right)} + \sigma_y\sin{\left(\phi\right)} \right) ~\theta\left(1+\epsilon\right)/2} \label{roteq}
\eeq
where $\sigma_x$ and $\sigma_y$ represent the
Pauli spin operators. 

The BB1 corrective sequence has the form:
\beq {\cal R}_0[\theta(1+\epsilon)]~{\cal R}_{\phi_1}[\pi(1+\epsilon)]~{\cal R}_{\phi_2}[2\pi(1+\epsilon)]~{\cal R}_{\phi_1}[\pi(1+\epsilon)]\label{bb1eq}.\eeq
The fidelity of this composite pulse can then be expanded in $\epsilon$; all orders of the expansion
up to and including coefficients of $\epsilon^5$ equal zero for
\beq
\phi_1=\arccos{\left(-\frac{\theta}{4\pi}\right)},~~\phi_2=3\phi_1.
\label{bb1phases}\eeq

For example, to achieve high-fidelity $\pi$ rotations, we choose
$\phi_1=0.580\pi$ (104.5$^{\circ}$) and $\phi_2=1.741\pi$ (313.4$^{\circ}$). Setting phases accurately is more difficult in EPR than in NMR, owing to the higher frequency; the fidelity of a
BB1 $\pi$ pulse depends on \emph{small} imperfections in the
pulse phases, $\delta\phi_1$ and $\delta\phi_2$, approximately as
\begin{eqnarray}\label{phaseeq}
1~-&(0.75 \delta\phi_1^2 - 1.125 \delta\phi_1\delta\phi_2 + 0.5
\delta\phi_2^2)~\epsilon^2\pi^2,
\nonumber\\
&{-}\: (0.121\delta\phi_1 - 0.091
\delta\phi_2)~\epsilon^4\pi^4+O(\epsilon^6) .
\end{eqnarray}

Qubit rotation is achieved in EPR through an on-resonance
microwave pulse of controlled power and duration. Rotation angle
errors therefore arise from either pulse duration errors (which
can be assumed uniform throughout the sample), or errors in the
magnitude of the microwave field, $B_1$, which varies across the sample depending on
the homogeneity of the EPR cavity mode. For single-molecule
manipulation there is no inhomogeneity, but
limitations in the resolution of power would
still lead to systematic errors.

\begin{figure}[t]
\centerline {\includegraphics[width=3.6in]{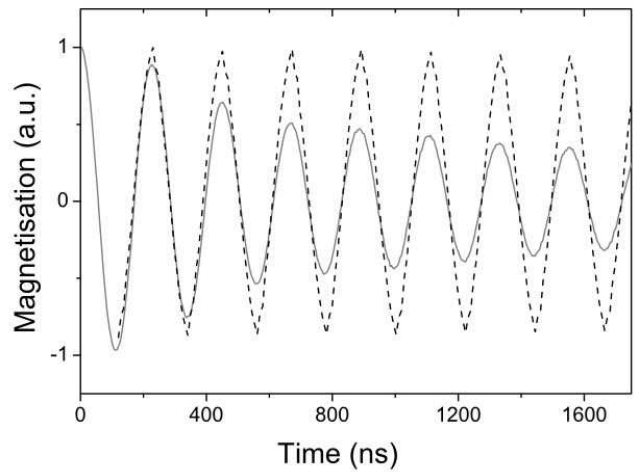}}
\caption{Rabi oscillations for \emph{i}-NC$_{60}$ in CS$_2$ at 190
K (solid curve). \emph{BB1-Rabi} oscillations exploiting BB1
composite pulses to remove the decay caused by pulsed field
inhomogeneity (dashed curve).} \label{rabicomp}
\end{figure}

The effect of $B_1$ inhomogeneity is observed in the decay
of Rabi oscillations as the RF pulse duration increases (see Fig.~\ref{rabicomp}, solid curve). The inhomogeneity in $B_1$ causes an inhomogeneity in the Rabi period, so spins in different parts of the sample gradually lose coherence under the influence of the RF pulse. By removing the error accumulated over these
long rotations, BB1 composite pulses can be used to obtain Rabi
oscillations that would be observed in the absence of an inhomogeneous RF field.

For every desired rotation angle there are two distinct phases
which are required in the correction sequence (Eqn.~3). In order to be able
to use the same two phases over the course of the experiment, the
long RF pulses were divided into separate high-fidelity $\pi$
pulses, with the remainder provided by a simple pulse of length
$\theta$, where $0 < \theta < \pi$. The \emph{BB1-Rabi}
oscillations shown in the dashed curve in Fig.~\ref{rabicomp} were obtained with the
pulse sequence
\beq {\cal R}_{0}[\theta]-\left(~{\cal R}_{0}[\pi]-{\cal
R}_{0.58\pi}[\pi]-{\cal R}_{1.74\pi}[2\pi]-{\cal R}_{0.58\pi}[\pi]~\right)_n ,
\eeq
and demonstrate that this sequence does not accumulate errors arising from a $B_1$ inhomogeneity.
Since $T_2\approx200~
\mu$s, which is long compared with one Rabi time period, electron spin decoherence is small on the timescale shown in Fig.~\ref{rabicomp}. 

\begin{figure}[t]
\centerline {\includegraphics[width=3.6in]{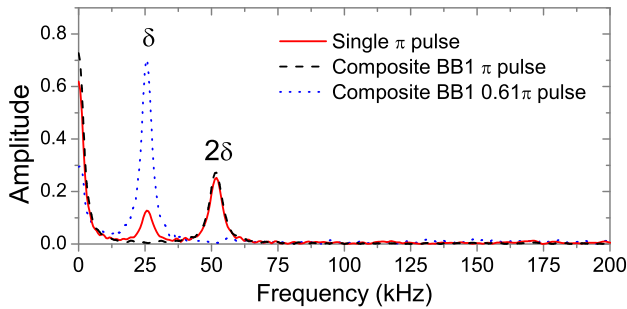}}
\caption{Fourier transform of ESEEM in \natc~shows two frequency
components 26 and 52 kHz. For simple $\pi$ refocusing pulses, the 26 kHz component is attributed to pulse error (solid line) and is removed
when a BB1 composite $\pi$ pulse is used (dashed line). For a refocusing pulse
of approximately $0.61\pi$, the 52~kHz component, which would be present due to
pulse error, can also be removed by using a BB1 composite $0.61\pi$ pulse (dotted line).} \label{eseemfig}
\end{figure}

In a recently reported mechanism for ESEEM, the frequency content of the echo modulation was found to have a clear dependence on the fidelity of the echo-refocusing
pulse~\cite{eseem03}. The echo decay is modulated by two
frequency components, $\delta$ and $2\delta$, whose relative
amplitudes are a function of the rotation angle of the 
refocusing pulse. For a $\pi(1+\epsilon)=\pi+\theta_{\epsilon}$ refocusing pulse, where $\theta_{\epsilon}$ represents the (small) absolute error in the refocusing pulse, the
relative magnitudes of the two components is given by 
\beq
\frac{F(\delta)}{F(2\delta)}=2\theta_{\epsilon}^2. \eeq 
However,
for a refocusing pulse of twice the magic angle~\cite{spinchoreo}: \beq \theta=
2\cos^{-1}{\left(\sqrt{1/3}\right)} +\theta_{\epsilon},\eeq the $\delta$ component
dominates:\beq
\frac{F(\delta)}{F(2\delta)}=\frac{\sqrt{2}}{\theta_{\epsilon}}.\label{eseemeq}\eeq Fig.~\ref{eseemfig} compares the Fourier Transform of the
echo decay using either simple or BB1 composite pulses, for
nominal refocusing pulses $\theta=\pi$ and $\theta=0.608\pi$. For the
latter rotation angle, phases $\phi_1=0.549\pi$ and $\phi_2=1.646\pi$ were used in the BB1 pulse. When the
error-compensated pulse is applied, the secondary frequency component
is removed in each case, further demonstrating the ability of this
composite pulse to correct for systematic errors in rotation
angle. It is worth noting that in these ESEEM experiments, the BB1 ``refocusing'' pulses operate on spins which are dispersed in the rotating x-y plane. This illustrates the effectiveness of the BB1 composite
pulse over a range of initial states.

\begin{figure}[t]
\centerline {\includegraphics[width=3.4in]{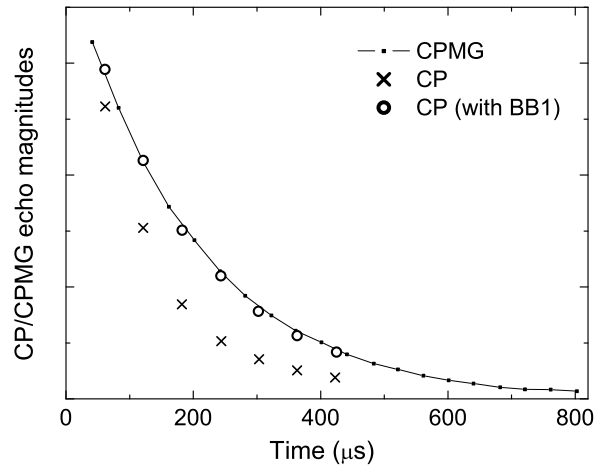}}
\caption{Comparison of the echo signal decays in the reference
CPMG sequence (dots) with that of the error-sensitive CP (crosses) sequence
provides a measure of rotation angle errors. CP echoes generated
with simple $\pi$ pulses decay more rapidly owing to rotation angle
errors. When the simple $\pi$ pulses are replaced with BB1
composite $\pi$ pulses (open circles), the decay rate is very close to
that in CPMG. } \label{cpcpmgfig}
\label{cp-cpmg}
\end{figure}

Rotation angle errors can be measured by comparing the rates of
echo decay in two multi-pulse sequences: Carr-Purcell (CP) and
Carr-Purcell-Meiboom-Gill (CPMG)~\cite{spinchoreo,morton04}. The
echo decay in CPMG shows no sensitivity to rotation angle errors
(after every even numbered cycle) and is dictated purely by
decoherence.  Echo magnitudes in CP show a cumulative sensitivity
to errors in the refocusing $\pi$ pulses and hence decay faster
than in CPMG, as shown in Fig.~\ref{cp-cpmg}. By performing a comparison as described in Ref.~\cite{morton04}, we estimated that the refocusing pulses had a standard error of $0.1\pi$ ($18^{\circ}$), equivalent to a fidelity of ${\cal F}=0.988$. 

By
replacing the $\pi$ pulses in CP with BB1 composite $\pi$ pulses we are
able to make an estimate of the residual rotation angle error in
the high-fidelity rotation. The error accumulated in the CP
sequence is dependent on the number of cycles applied and hence
the sensitivity of this technique is limited only by the maximum
number of pulses that can be applied in one experiment (with the spectrometer used for these experiments, this is currently
32, which implies fewer than 8 BB1 composite rotations). From the decay of the BB1 corrected CP sequence, we
are able to conclude that the rotation angle error is \emph{at most} $0.01\pi$ ($2^{\circ}$), corresponding to a fidelity ${\cal F}>0.9993)$. The phases used in this BB1 experiment were measured using a SPAM sequence~\cite{morton04} and
found to be $\phi_1=(0.587\pm0.008)\pi$ and $\phi_2=(1.742\pm0.011)\pi$. Using Eqn.~\ref{phaseeq} these imply
an expected fidelity ${\cal F}=0.9999$.

Error correcting codes will be crucial to the operation of a
quantum computer~\cite{steane03}, compensating for both decoherence errors and the
gate operation errors described here.  The threshold of error probability for these codes varies as a function of the overhead, but generally lies between $10^{-3}$ and $10^{-4}$,
or a fidelity greater than 0.999~\cite{cheng04}. Despite the inherent
10$\%$~systematic error, the composite pulses described here are
capable of producing operations which meet this threshold fidelity.
However, errors in pulse phase cause the fidelity to fall short of
the theoretical optimum of $1-10^{-6}$. In addition to the BB1
composite pulse described here, longer composite pulses have been
proposed which can arbitrarily reduce the sensitivity to
systematic error~\cite{chuang04}. In order for these higher-order
composite pulses to be effective in pulsed EPR, greater phase
control is required. An analysis of random errors present would
complement the current work in determining where the compromise
between lengthy pulse sequences and error reduction lies.

We would like to thank Wolfgang Harneit's group at the
Hahn-Meitner Institute for providing Nitrogen-doped fullerenes,
and John Dennis at QMUL, Martin Austwick
and Gavin Morley for the purification of \emph{i}-NC$_{60}$. We
also thank Jonathan Jones for stimulating and valuable
discussions, and the Oxford-Princeton Link fund for support on this project. 
This research is part of the QIP IRC www.qipirc.org (GR/S82176/01). GADB thanks EPSRC for a Professorial Research Fellowship (GR/S15808/01). AA is supported by the Royal Society.  Work at Princeton was supported by the NSF International Office through the Princeton MRSEC Grant No. DMR-0213706 and by the ARO and ARDA
under Contract No. DAAD19-02-1-0040.

\end{document}